\documentclass[12pt,preprint]{aastex}
\shorttitle{Kumar \& Gangadhara}  
\shortauthors{} 
\received{}         
\begin{document}    
\title{INFLUENCE OF THE POLAR CAP CURRENT ON PULSAR POLARIZATION}
\author{D. Kumar and R. T. Gangadhara}  
\affil{Indian Institute of Astrophysics, Bangalore -- 5600034, India\\
dinesh@iiap.res.in, ganga@iiap.res.in}  
\altaffiltext{}{}     
\begin{abstract} 
We have developed a model for the polarization of curvature radiation
by taking into account the polar-cap-current induced perturbation
on the dipolar magnetic field. We present the effects of the polar cap
current on the pulsar radio emission in an artificial case when the rotation effects, such as aberration and
retardation, are absent. Our model indicates that the intensity
components and the polarization angle inflection point can be shifted
to either the leading or the trailing side depending upon the prevailing
conditions in the viewing geometry, the non-uniformity in source
distribution (modulation), and the polar cap-current-induced
perturbation. Also, we find evidence for the origin of symmetric-type circular polarization in addition to the antisymmetric type. Our
model predicts a stronger trailing component compared to that on the leading side of a given cone under some specific conditions.
\end{abstract}     
\section{Introduction}
Pulsars are universally accepted as fast rotating neutron stars
with very a strong magnetic field ($\sim10^8$-$10^{12}$ G), which is
predominantly dipolar. Their radio emission is thought to be curvature
emission from relativistic plasma streaming out along the open field
lines of a dipole whose foot points on the neutron star surface define
the polar cap \citep{S71,RS75,CR80}. The simplistic picture of pulsar
radio emission that we can gather from the polar cap emission models
is as follows. The rotationally induced, very strong electric field
component parallel to the magnetic field pulls out the primary plasma
from the neutron star surface and accelerates it to very high
energies with a Lorentz factor of $\gamma\sim 10^5$-$10^7.$ These
ultrarelativistic primary particles emit $\gamma$-ray photons through
curvature and synchrotron radiations, which in turn produce the secondary
electron-positron ($e^{-}$, $e^{+}$) pair plasma by their interaction
with a strong magnetic field. The secondary plasma further undergoes
the pair cascade process and develops into a high density plasma of
$\gamma\sim 10^2-10^3.$ Once the density of pair plasma becomes
comparable to the Goldreich-Julian charge density, it shields the parallel
electric field component \citep{GJ69}. Hence, the plasma motion
becomes ``force free'' along the field lines and emits curvature
radiation due to the curvature of the field lines. Since the
brightness temperature of the pulsar radio emission is very high
($\sim10^{25}$-$10^{30}$ K), the emission is believed to be coherent
emission from plasma.

Pulsars show a high degree of linear polarization with a systematic ``S''-
shaped polarization position angle (PPA) swing, which is a
characteristic property of pulsar signals. The
``rotating-vector-model'' (RVM) of \citet{RC69} attributes this
characteristic ``S'' curve to an underlying geometry, wherein the magnetic
field is assumed to be mainly dipolar, and relativistic beaming is in
the direction of field line tangents. Ever since the success of RVM,
the fitting of pulsar polarization angle profiles to the model has been
attempted to constrain the underlying emission geometry, such as the
magnetic axis inclination angle with respect to the rotation axis and the
sight line impact angle with respect to the magnetic axis. However, some
pulsars show that the polarization angle behavior deviates from the
standard ``S'' curve, particularly in millisecond pulsars, where the
polarization sweep is noisy and on average flatter
\citep{Xetal98}. Several relativistic and plasma effects have been
proposed to understand these deviations: plasma propagation effects
\citep{BA86, Mc97}, aberration of the beaming direction from strict
parallelism \citep{BCW91,D08,KG12}, distortion of the underlying dipole
field due to the field aligned polar cap current \citep[PC-current][]{HA01}, or multiple interacting orthogonal polarization modes
\citep{McS98}.

Phenomenologically, pulsar emission is recognized as a central ``core''
arising from a region near the magnetic pole and surrounding
``cones'' from concentric rings around the pole
\citep[e.g.,][]{R83,R90,R93,MD99,GG01,MR02}. However, due to an
asymmetry in the phase location of the components, there are some contrary
arguments that the emission is ``patchy'' \citep{LM88}. Pulsars also
show an asymmetry in the strength of the components intensity between the
leading and trailing sides.  The asymmetry observed both in the
strength and the phase location of the components has been attributed to
the effects of rotation such as aberration and retardation
\citep[e.g.,][]{BCW91,GG01,GG03,DRH04,TG07,D08,DWD10,Tetal10,KG12}. Due
to the rotationally induced aberration and asymmetry in the curvature
of the source trajectory, the inflection point of the PPA profile lags the
peak of the central intensity component (core)
\citep{BCW91,TG10,D08,KG12}, and the leading side components become
stronger and broader than the corresponding ones on the trailing side
\citep{TG07,DWD10,Tetal10,KG12}.

In general, circular polarization is common in pulsar emission, but
diverse in nature \citep{Hanetal98,YH06}. The research related to
its origin is still of great importance from the point of view of the
emission mechanism. There are two types of models for the origin of the
circular polarization: intrinsic to the emission mechanism
\citep[e.g.,][]{M87,GS90a,GS90b,RR90,G97,G10,KG12} and generated by
the propagation effects \citep[e.g.,][]{CR79,M03}. Only the
``antisymmetric''-type circular polarization was thought to be an
intrinsic property of curvature radiation. Very recently, \citet{KG12}
have shown that the ``symmetric''-type circular polarization (where
the circular sign remains the same throughout a component or pulse)
is also possible within the framework of the curvature radiation, which
takes into account the pulsar rotation and modulation (nonuniform
plasma distribution). They have made an attempt to simulate the
diverse behavior of polarization properties of pulsar radio emission
by developing a relativistic model on full polarization of curvature
radiation.

Since the polarization behavior of pulsars is believed to depend upon
the underlying geometry of the emission region, any perturbation to the
dipole field is expected to be reflected in the polarization
profiles. The induced toroidal magnetic field due to field-aligned,
poloidal currents above the polar cap perturbs the dipole field and it
is quite significant in affecting the PPA \citep{HA01} and introduces a 
phase shift into the intensity components \citep{G05}. It is shown
that the PC-current can only introduce an offset into the PPA swing, but
there is no significant phase shift into the inflection point. The full
polarization model that describes all the polarization parameters of
radio emission in the PC-current perturbed dipole magnetic field does
not exist. So, in an attempt to understand the PC-current-induced
perturbation on the pulsar radio emission and polarization, we develop
a complete polarization model by taking into account only the
PC-currents. We derive the viewing geometry for the perturbed dipole
field in Section 2. In Section 3, we present the simulation of typical
polarization profiles for the cases of uniform and nonuniform
distribution of radiation sources. In Section 4 we provide the
discussion, and in Section 5 the conclusion.
\section{PERTURBATION OF THE MAGNETIC FIELD AND EMISSION GEOMETRY}
In spherical polar coordinates $(r,~\theta,~\phi)$ centered on the
magnetic axis, the unperturbed dipole field is given as
\begin{equation}
 \mbox{\boldmath $B_0$}=\left(\frac{2 \mu \cos\theta}{r^3},~\frac{\mu
   \sin\theta}{r^3},~0\right),
\end{equation}
where $\mu$ is the magnetic moment, $\theta$ is the magnetic
colatitude, and $r$ is the radial distance from the origin. Most of the
polar cap models of pulsar emission include a current of relativistic
charged particles streaming out along the open field lines, which is
approximately equal to the Goldreich--Julian charge density ($J_{\rm GJ}$),
\begin{equation}
 \mbox{\boldmath$J$}=-\frac{1}{2 \pi}~\varsigma~(\mbox{\boldmath
   $\Omega$}\cdot\mbox{\boldmath$B_0$})\hat B_0~,
\end{equation}
where $\varsigma=J/J_{\rm GJ}$ is a scale factor of order unity,
reflecting our ignorance of the actual current, and $\mbox{\boldmath
  $\Omega$}$ is the pulsar angular velocity. For
simplicity, we treat $\varsigma$ to be a constant quantity
throughout this work.

By assuming a small polar cap so that the variation in
$\mbox{\boldmath $\Omega$}\cdot\mbox{\boldmath$B_0$}$ across the polar
cap is also small, and that $\mbox{\boldmath$B_0$}$ is a point dipole
with an axisymmetric current flow, the induced magnetic field due to the
field aligned current is given as \citep{HA01}
\begin{equation}
 \mbox{\boldmath $B_1$}=\left(0,~0,~-\frac{2 \mu
   \sin\theta}{r^3}~\varsigma\frac{r}{r_{\rm LC}} \cos\alpha\right),
\end{equation}
which is purely toroidal and axisymmetric with respect to the magnetic
axis, i.e., independent of the magnetic azimuth $\phi$. The angle
$\alpha$ is the magnetic axis inclination angle, and $r_{\rm
  LC}=c/\Omega=c P/(2 \pi)$ is the light cylinder radius, where $c$ is
the speed of light and $P$ is the pulsar rotation period. Then the
perturbed dipole field $\mbox{\boldmath $B$}=\mbox{\boldmath
  $B_0$}+\mbox{\boldmath $B_1$}$ with $B_{1}\ll B_{0}$ is also
axisymmetric with respect to the magnetic axis.

The differential equations for the magnetic field lines are
\begin{equation}
 \frac{dr}{B_r}=\frac{r~d\theta}{B_\theta}=\frac{r \sin\theta~ d\phi}{B_\phi}~,
\label{eqn:diff}
\end{equation}
where $B_r,~B_\theta,~\rm{and}~B_\phi$ are the respective
$r,~\theta,~\rm{and}~\phi$ components of $\mbox{\boldmath $B$}$ in
polar coordinates.  Substituting the expressions for
$B_r,~B_\theta,~\rm{and}~B_\phi$ into Equation (\ref{eqn:diff}) yields
the following three differential equations for the magnetic field
lines:
\begin{equation}
 \label{eqn:diff1}
 \frac{1}{r}~dr= 2 \cot\theta~d\theta,
\end{equation}
\begin{equation}
 \label{eqn:diff2}
 \frac{r}{\sin\theta}~d\theta=-\frac{r_{\rm LC}}{2\varsigma \cos\alpha}~d\phi,
\end{equation}
and
\begin{equation}
 \label{eqn:diff3}
 \frac{1}{\cos\theta}~dr=-\frac{r_{\rm LC}}{\varsigma \cos\alpha}~d\phi.
\end{equation}
Since Equation (\ref{eqn:diff3}) can be deduced from Equations
(\ref{eqn:diff1}) and (\ref{eqn:diff2}), magnetic field lines can be
defined uniquely by the earlier two equations.

\begin{figure}
\centering
\epsscale{1.2}
\plotone{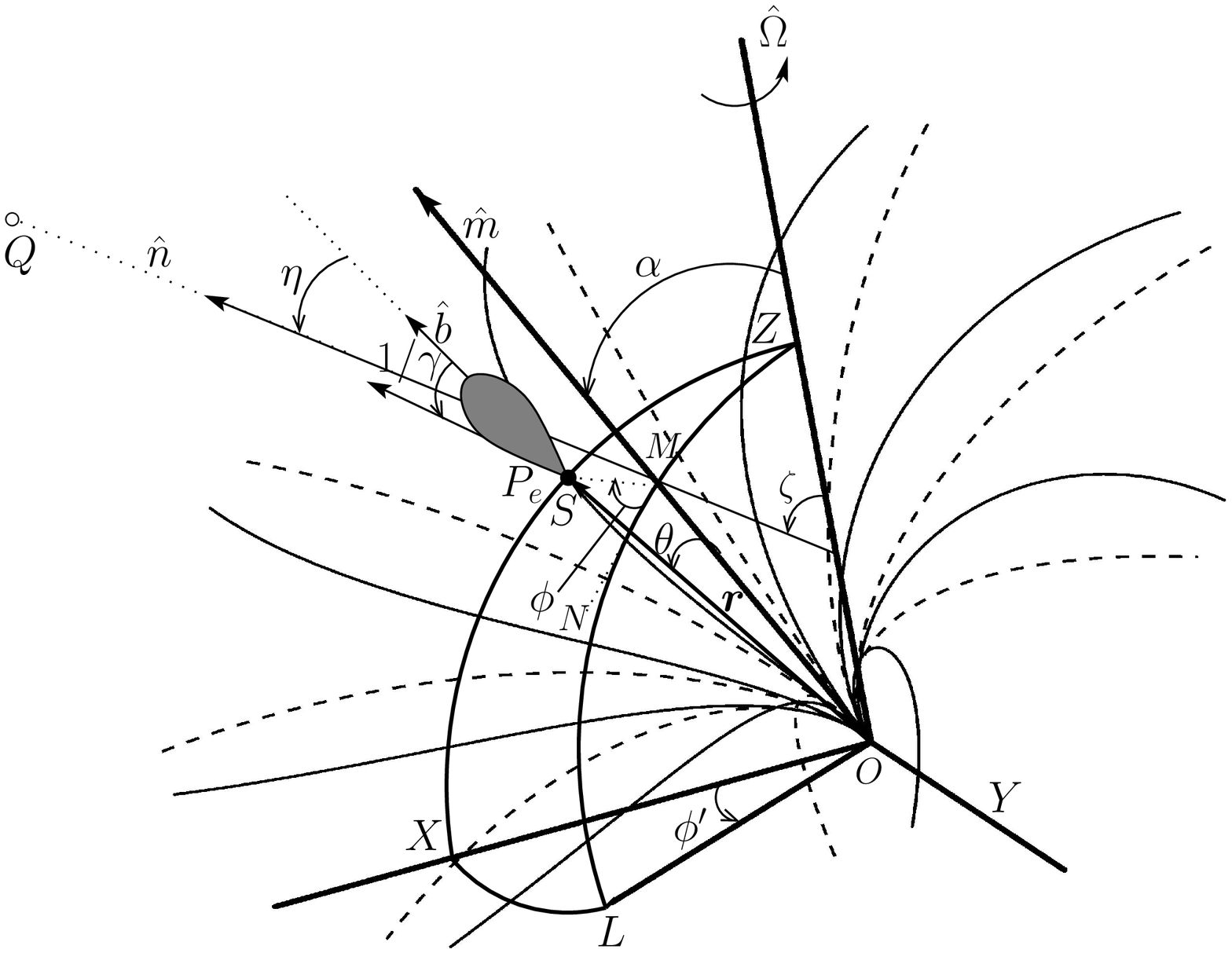}
\caption{Geometry for curvature radiation from a relativistic source S
  in a distorted (or perturbed) magnetic dipole due to polar cap
  currents. Magnetic axis $\hat{m}$ is inclined by an angle $\alpha$
  with respect to pulsar rotation axis $\hat{\Omega}.$ The dashed line
  curves represent the field lines of a pure dipole whereas the solid
  line curves represent those of a perturbed dipole. The observer's
  sight line $\hat{n}$ lies in the fiducial plane ($XZ$-plane), and the
  observation point $Q$ is at a distance $R$ from the emission point $\rm
  P_e$. We have used $\alpha=30^\circ$, $\phi'=30^\circ$,
  $r_{e}=1~r_{\rm LC}$, $P=1$ s, $\varsigma=1$, $\phi_{i}$ from
  $0^\circ$ to $360^\circ$ in steps of $45^\circ$,
  $\theta_{i}=0^\circ$, and $r$ from $0$ to $0.6~r_{\rm LC}$ to sketch
  the field lines of the pure dipole and those of the perturbed
  dipole.}
 \label{fig:Figure1}
\end{figure}

Integration of Equation (\ref{eqn:diff1}) gives 
\begin{equation}
\label{eqn:r1}
 r=K_1 \sin^{2}\theta,
\end{equation}
where $K_1$ is the integration constant. Using $r=r_{\rm max}=r_e$ the
field line constant at $\theta=\pi/2,$ we obtain $K_1=r_e.$ Then
Equation (\ref{eqn:r1}) becomes
\begin{equation}
\label{eqn:r2}
 r=r_{e} \sin^{2}\theta.
\end{equation}
Substitution of Equation (\ref{eqn:r2}) into Equation
(\ref{eqn:diff2}) and integration gives
\begin{equation}
\label{eqn:phi1}
 r_{e} \cos\theta=\frac{r_{\rm LC}}{2\varsigma \cos\alpha}~\phi+K_{2}~,
\end{equation}
where $K_{2}$ is the integration constant. Using $\phi=\phi_{i}$ at
$\theta=\theta_{i},$ we obtain $K_{2}= r_{e}\cos\theta_{i}-r_{\rm
  LC}~\phi_{i}/(2\varsigma \cos\alpha)$, where $\theta_{i}$ and
$\phi_{i}$ are the initial colatitude and azimuth, respectively.
By substituting  $K_{2}$ into Equation (\ref{eqn:phi1}) we get
\begin{equation}
 \phi=\phi_{i}+\frac{2 r_{e}}{r_{\rm LC}}~\varsigma
 \cos\alpha~(\cos\theta-\cos\theta_{i}).
\label{eqn:phi}
\end{equation}
The position vector $\mbox{\boldmath $r$}$ of an arbitrary point $P_e$
on a perturbed dipolar field line is given by Equation (1) of
\citet{G10}, where the parameter $\phi$ has to be replaced by the
expression that we have given in Equation (\ref{eqn:phi}). By using
$\alpha=30^\circ,$ rotation phase $\phi'=30^\circ,$
$r_{e}=r_{\rm LC},$ $P=1$~s, $\varsigma=1,$ $\theta_{i}=0^\circ,$
$\phi_{i}$ from $0^\circ$ to $360^\circ$ in steps of $45^\circ$, and
$r$ from $0$ to $0.6~r_{\rm LC}$ we have traced the field lines and
presented them in Figure \ref{fig:Figure1}. Dashed line curves represent
the field lines of an actual dipole while the solid line curves
represent those of the perturbed dipole. PC-currents induce the
toroidal component of the magnetic field, and hence there is an azimuthal
twist in the field lines of the perturbed field with respect to the
magnetic axis.

The curvature of the perturbed dipolar field line is given by
\begin{equation}
 \mbox{\boldmath
   $k$}=\frac{d\hat{b}}{ds}=\frac{1}{|\textbf{b}|}\frac{d\hat{b}}{d\theta}~,
\label{eqn:k}
\end{equation}
where $ds=|\textbf{b}| d\theta$ is the arc length of the field line,
$\hat{b}=\mbox{\boldmath $b$}/|\mbox{\boldmath $b$}|,$ and
$\mbox{\boldmath $b$}=d\mbox{\boldmath $r$}/d\theta$ is the field line
tangent. The magnitude of $\mbox{\boldmath $b$}$ can readily found
to be
\begin{equation}
 |\mbox{\boldmath $b$}|=\frac{r_{e}\sin\theta}{2}
 \sqrt{10+6\cos2\theta+4\Delta ^2\sin^2\theta},
\end{equation}
where the parameter $\Delta=2 \varsigma r \cos\alpha/r_{\rm LC}.$
Then, the radius of the curvature $\rho=1/|\mbox{\boldmath $k$}|$ of a field
line becomes
\begin{equation}
 \rho=\frac{ r_{e} \sin \theta ~d_1^2}{2 \sqrt{2} \sqrt{D_1 + D_2 \cos
     2\theta - D_3 \sin 2 \theta}},
\label{eqn:rho}
\end{equation}
where $\mbox{\boldmath $k$}=d\hat{b}/ds$ is the curvature vector of
the field line,
\begin{eqnarray}
 D_1 & = & 2 \left(9+18 \Delta ^2+\Delta ^4\right)
 d_1^2+\left(5+\Delta ^2\right) d_2^2, \nonumber \\ D_2 & = & 24
 \Delta ^2 d_1^2+\left(3-\Delta ^2\right) d_2^2, \nonumber \\ D_3 & =
 & 6 \left(1-\Delta ^2\right) d_1 d_2, \nonumber
\end{eqnarray}
and 
\begin{eqnarray}
 d_1 & = & 10 + 6 \cos 2 \theta + 4 \Delta ^2 \sin ^2\theta, \nonumber \\
 d_2 & = & 6 \left(1-\Delta ^2\right) \sin 2 \theta. \nonumber
\end{eqnarray}
The velocity of the relativistic source (particle or plasma bunch), which
is constrained to move along the perturbed field line, is given by
\begin{equation}
 \mbox{\boldmath $v$}=\kappa~c~\hat{b}~,
\label{eqn:v}
\end{equation}
where the parameter $\kappa=\sqrt{1-1/\gamma^{2}}$ with $\gamma$ being
the source Lorentz factor.
Then the acceleration $\mbox{\boldmath $a$}=d\mbox{\boldmath $v$}/dt$
due to the curvature of field lines is given by
\begin{equation}
 \mbox{\boldmath $a$}=(\kappa c)^{2} \mbox{\boldmath $k$}~,
\label{eqn:a}
\end{equation}
where we have used $ds=|\textbf{b}| d\theta=\kappa c~ dt.$

The accelerating relativistic particles emit beamed emissions in the
direction of their velocities. At any given rotation phase
$\phi'$ and emission altitude $r$, the observer receives the
emissions from a finite beaming region consisting of the flux of
magnetic field lines. The observer receives a maximum beamed emission when
sight line $\hat{n}$ exactly aligns with the source velocity
$\mbox{\boldmath $v$}.$ Observer sight line
$\hat{n}=(\sin\zeta,~0,~\cos\zeta),$ where the angle
$\zeta=\alpha+\sigma$ with $\sigma$ being the sight line impact angle
with respect to the magnetic axis. Since the beaming angle of the beamed
emission from a relativistic particle is $\sim1/\gamma$ about its
velocity, the beaming region boundary can be specified by the emission
points at which $\hat n \cdot \hat b=\cos (1/\gamma).$ Therefore the
emission points within the beaming region satisfy the condition
$-1/\gamma \leq \eta \leq 1/\gamma$, where $\eta$ is the angle between
$\hat n$ and $\hat b$. Following Kumar \& Gangadhara (2012), we define
the emission point coordinates: magnetic colatitude and azimuth of the
emission points within the beaming region as $\theta_{e}$ and
$\phi_{e},$ respectively, wherein at the center of the beaming region
$\theta_{e}=\theta_{0}$ and $\phi_{e}=\phi_{0}$.

Next, for a given emission altitude $r$, we find the emission point
coordinates $(\theta_{0},\, \phi_{0})$ 
and the range of $\theta$ and $\phi,$ which define the beaming region
boundary. Solving $\hat{m}\cdot \hat{b}=\cos\tau=(1+3 \cos 2 \theta)
/\sqrt{10+6 \cos 2 \theta+4 \Delta ^2 \sin ^2\theta}$ for $\theta$
gives
\begin{equation} 
\cos2\theta=\frac{1}{3} \left(\left(1-\frac{\Delta ^2}{3}\right) \cos
   ^2\tau +\cos \tau  \sqrt{\left(1-\frac{\Delta
   ^2}{3}\right)^2 \cos ^2\tau +8 \left(1+\frac{\Delta
   ^2}{3}\right)}-1\right),
\label{eqn:theta}
\end{equation}
where $\tau$ is the angle between $\hat{m}$ and $\hat{b}$. But the
angle $\Gamma$ between $\hat{m}$ and $\hat{n}$ is given by
\begin{eqnarray}
 \cos\Gamma=\cos \alpha 
   \cos \zeta +\sin \alpha  \sin \zeta  \cos \phi' \nonumber.
\end{eqnarray}
For an exact alignment of $\hat{n}$ and $\hat{b}$ at the beaming
region center, $\hat{n}\cdot \hat{b}=1,$ which implies $\tau=\Gamma.$
Hence we have
\begin{equation} 
\cos2\theta_{0}=\frac{1}{3} \left(\left(1-\frac{\Delta ^2}{3}\right) \cos
   ^2\Gamma +\cos \Gamma  \sqrt{\left(1-\frac{\Delta
   ^2}{3}\right)^2 \cos ^2\Gamma +8 \left(1+\frac{\Delta
   ^2}{3}\right)}-1\right).
\label{eqn:theta0}
\end{equation}
Next, by solving $\hat n \times \hat b=0$, we obtain
\begin{equation} 
\sin\phi_{0}=\frac{\csc \Gamma (\sin \zeta (\Delta \cos \alpha \cos
  \phi'-3 \cos \theta \sin \phi')-\Delta \sin \alpha \cos \zeta
  )}{\sqrt{\Delta ^2+9 \cos ^2\theta }}
\label{eqn:sinphi0}
\end{equation}
and
\begin{equation}
\cos\phi_{0}=\frac{\csc \Gamma (\sin \zeta (3 \cos \alpha \cos \theta
  \cos \phi'+\Delta \sin \phi')-3 \sin \alpha \cos \zeta \cos \theta
  )}{\sqrt{\Delta ^2+9 \cos ^2\theta }}.
\label{eqn:cosphi0}
\end{equation}
In Figure \ref{fig:Figure2} we have plotted the emission 
coordinates $(\theta_{0}, \,\phi_{0})$ of the beaming region center
and the corresponding radius of curvature $\rho$ of the source trajectory at
those points as functions of the rotation phase. We chose the parameters $\alpha=30^\circ$,
$\sigma=\pm 5^\circ$, the emission altitude $r_{n}=r/r_{\rm LC}=0.1$,
$P=1$ s, and $\varsigma=1$. The dotted line curves are for the
parameters in the case of an unperturbed dipole, and the solid line curves
represent those in the perturbed case. The colatitude $\theta_{0}$ is
found to remain unaffected, whereas there is an upward vertical
shift in azimuth $\phi_{0}$ due to the induced toroidal magnetic
field. Therefore at any rotation phase the radiation 
aberrated in the direction opposite to the direction of the pulsar
rotation for positive $\sigma$ (i.e., for $\zeta>\alpha$), whereas for
negative $\sigma$ (i.e., for $\zeta<\alpha$) it is aberrated in the
direction of the pulsar rotation. In other words, emission locations in
$\phi$ get shifted to later phases with respect to those due to
the unperturbed dipole for positive $\sigma$, whereas for negative
$\sigma$ they get shifted to earlier phases (the phase shift of an
antisymmetric point of $\phi_0$ with respect to $\phi'=0$ is
indicated by the longer arrows). Note that the inflection point of
$\phi_0$ (the phase at which the $\phi_0$ curve has a maximum slope) is found to
remain unaffected as indicated by the shorter arrows. Since the effect of the
induced toroidal component of the magnetic field due to PC-currents is just a twist of the field lines around the magnetic axis and
is the same on both the leading and trailing sides, $\rho$ in the perturbed
configuration will remain almost symmetric with respect to the phase
$\phi'=0^\circ,$ similar to that in the unperturbed one but with
smaller values because of an induced curvature.

Next at $\phi=\phi_{0}$, we solve $\hat n \cdot \hat
b=\cos\eta_{\rm max}=\cos (1/\gamma)$ and find the allowed range of
$\tau$: $\Gamma -1/\gamma\leq\tau\leq\Gamma +1/\gamma,$ and hence
deduce the allowed range of $\theta$ from Equation
(\ref{eqn:theta}). Then for any $\theta$ within its allowed range, we
find the range of $\phi$: $\phi_{0}-\delta \phi \leq \phi \leq
\phi_{0}+\delta \phi$ by solving $\hat n \cdot \hat b=\cos 1/\gamma.$
Note that the expression obtained for $\delta \phi$ is the same as that
given in \citet{G10} except for the fact that the angle $\tau$ has to
be replaced with the one that we have given above. We note that the
behavior of the beaming regions with respect to rotation phase $\phi'$
is mostly similar to the case of the unperturbed dipole of the non-rotating
magnetosphere \citep{G10}. Note that the whole beaming regions of the
perturbed dipole get shifted in the coordinate $\phi$ with respect to
those of the unperturbed dipole (see Figure \ref{fig:Figure2}).
\section{POLARIZATION STATE OF THE RADIATION FIELD}
The spectral distribution of the radiation field emitted by 
accelerating relativistic plasma at the observation point $Q$ is given
by \citep{Jackson98}
 \begin{equation}
\textbf{E}(\textbf{r},\omega) = \frac{1}{\sqrt{2 \pi}}\frac{qe^{i \omega
    R_{0}/c}}{R_{0}~c}\int^{+\infty}_{-\infty}
\frac{|\textbf{b}|}{\kappa c}\frac{\hat{n}\times[(\hat{n} -
    \mbox{\boldmath $\beta$})\times\mbox{\boldmath
      $\dot{\beta}$}]}{\xi^{2}} e^{i \omega (
  t-\hat{n}\cdot \textbf{r}/c)} d\theta.
\label{eqn:Ew} 
\end{equation} 
Note that the time $t$ in the above equation has to be replaced by the
expression given in Equation (6) of Kumar \& Gangadhara
(2012). Following the method given in Kumar \& Gangadhara (2012), we
solve the integral of Equation (\ref{eqn:Ew}) and estimate the
polarization state of the radiation field in terms of the Stokes
parameters $I,$ $Q,$ $U,$ and $V,$ which are defined in \citet{G10}.
\subsection{Emission with Uniform Radiation Source Density}
By assuming a uniform radiation source distribution in the emission
region of the magnetosphere, we estimated the polarization state of the
emitted radiation. The simulated contour plots of total intensity $I$,
linear polarization $L=\sqrt{Q^2+U^2}$, and circular polarization $V$
in the perturbed dipole configuration are given in Figure
\ref{fig:Figure3}. The parameters used for simulation are
$\alpha=30^\circ$, $\sigma=\pm 5^\circ$, $r_{n}=0.1$, $P=1$ s,
$\varsigma=1$, $\gamma=400$, $\nu=600,$ MHz and the rotation phase is
$\phi'=0^\circ$. Note that the parameters in each panel are
normalized with the corresponding peak value of the total intensity. Also
note that the contour plots are plotted in such a way that the
emissions go from the leading side to the trailing side of the beaming
regions in azimuth $\phi$.

The polarized emissions from the beaming region get rotated as
indicated by the contour patterns in the $(\theta,\phi)$-plane compared
to those due to the unperturbed dipole \citep[see Figure 3
  of][]{KG12}. The rotation of the contour patterns is due to the
perturbation of the actual dipole geometry by the PC-currents. Also
it introduces an asymmetry into the beaming region emissions in such a
way that there is a selective enhancement of emissions over larger
the $\theta$ and larger $\phi$ regions due to a larger resultant
curvature. This is very clear from the $V$ parameter contour plots
where there is a selective enhancement of the positive circular over the
negative circular. Note that the direction of the rotation of contour
patterns in the $(\theta,\phi)$-plane is opposite between the positive
and negative $\sigma$ cases. It is in opposite direction to the one with a
rotating dipole for positive $\sigma$ while for negative $\sigma$ it
is in the same direction \citep[see][]{KG12}. Also note that the magnitude
of the rotation of intensity patterns is found to be independent of
the rotation phase of the magnetic axis unlike the one other due to the effect of the
pulsar rotation where it noticeably decreases at outer phases.

The net emission that the observer receives at any instant of time
will be an incoherent superposition of emissions from different plasma
particles/bunches. Following \cite{G10}, we have simulated the
polarization profiles: total intensity $I_s,$ linear polarization
$L_s,$ circular polarization $V_s,$ and polarization position angle
$\psi_s=(1/2)\tan^{-1}(U_s/Q_s),$ and plotted them in Figure
\ref{fig:Figure4}. For simulation, we used the same parameters as
in Figure \ref{fig:Figure3} and assumed that the source distribution
is uniform throughout the emission region. The emission on the leading
side ($\phi'<0^\circ$) is roughly the same as that on the railing side
($\phi'>0^\circ$) due to the symmetry of curvature of field lines
on both the sides (see Figure \ref{fig:Figure2}), which is the same as in
the unperturbed dipole geometry. Even though emissions from individual
sources are highly polarized, the net emission from the whole beaming
region becomes less polarized due to the incoherent addition. However
a small quantity of positive circular polarization survives due to the
asymmetry in the magnitudes of positive and negative polarities (see
Figure \ref{fig:Figure3}) which becomes more significant when the
source distribution becomes nonuniform. The PPA curves (solid line curves) shift upward with respect to those
due to the unperturbed RVM model (dotted line curves) only, but the PPA
inflection point is left mostly unchanged. For comparison, PPAs derived by
\citet{HA01} are also superposed with our simulations and both are in
good agreement (see the dashed line curves). Note that \citet{HA01}
considered the emission from the beaming region center only, and the
simulations show that the contributions from the neighboring emission
points will have the least effect on the net PPA. However they become
important once there is a density gradient in the plasma distribution
as we show later.
\subsection{Radiation Emission from the Nonuniform Distribution of Sources}
The sub-pulses in the pulsar individual pulses suggest that the emission
region of the pulsar magnetosphere may not be filled with a uniform plasma
density. When the observer's sight line encounters plasma
columns which are associated with the sparks activity on the polar cap,
components such as features in intensity can result. Hence the net
emission that the observer receives will be modulated emission over
the uniform background emission. We model the nonuniform source
distribution in the emission region in terms of ``modulation''
function as considered in Kumar \& Gangadhara (2012). The intensity
components of average profiles are assumed to be nearly Gaussian in
shape, and hence they have been fitted with an appropriate Gaussian
\citep[e.g.,][]{Ketal94}. We define the modulation function as
\begin{equation}
f(\theta,\phi) = f_0~f_\theta~f_\phi,
\label{eqn:mod}
\end{equation} 
where $f_{0}$ is the amplitude, and $f_\theta
=\exp\left[-(\theta-\theta_{p})^{2}/\sigma_{\theta}^{2}\right]$ and
$f_\phi= \exp\left[-(\phi-\phi_{p})^{2}/\sigma_{\phi}^{2}\right]$ are
the Gaussian modulations in $\theta$ and $\phi$, respectively. The
parameters $(\theta_{p}, \phi_{p})$ define the peak location of the
Gaussian and $\sigma_{\theta} = w_{\theta}/(2\sqrt{ln 2})$ and
$\sigma_{\phi} = w_{\phi}/(2\sqrt{ln 2})$, where
$w_{\theta}$ and $w_{\phi}$ are the corresponding FWHM of the Gaussian in the two directions.
\subsubsection{Emission with Azimuthal Modulation}
By considering a modulation in the magnetic azimuthal direction that has a 
peak location in the magnetic meridional plane in two cases
$(\phi_p=0^\circ$ for $\sigma=5^\circ$ and $\phi_p=180^\circ$ for
$\sigma=-5^\circ),$ we computed the polarization profiles 
 and shown them in Figures \ref{fig:Figure5} and
\ref{fig:Figure6}, respectively. For the simulation we used the parameters
$f_0=1$, $f_\theta=1$, and other parameters that are the same as in Figure
\ref{fig:Figure3}. In the case of $\sigma=5^\circ$ (see Figure
\ref{fig:Figure5}), the intensity peak shifted to a later phase 
 whereas for $\sigma=-5^\circ$ it gets shifted to an
earlier phase (see Figure \ref{fig:Figure6}). Due to the perturbation
of the magnetic field, the emission locations are shifted in the magnetic
azimuth as shown in Figure \ref{fig:Figure2}, and hence, the intensity gets
phase shifted to the later and earlier phases for the positive and
negative $\sigma$ sight lines, respectively. The phase shifts of the
intensity peaks in the case of $\sigma=5^\circ,$ $\sigma_\phi=0.1$ and
$0.4$ are found to be $0^\circ.54$ and $7^\circ.22,$ respectively,
whereas those in the case of $\sigma=-5^\circ,$ $\sigma_\phi=0.1$ and
$0.25$ are found to be $-0^\circ.75$ and $-1^\circ.50,$
respectively. The increase in the phase shift of intensity peaks with
increasing $\sigma_\phi$ can be explained as follows. In the perturbed
dipole geometry, the emission locations (particularly in the magnetic
azimuth) get phase shifted with respect to the unperturbed
one. Because of this phase shift, the curvature of the source
trajectory becomes asymmetric about the antisymmetric point of the
magnetic azimuth. For $\sigma=5^\circ$ the curvature becomes larger on the
trailing side of $\phi_p=0^\circ$ compared to that on the leading
side, whereas it is vice versa about $\phi_p=180^\circ$ for
$\sigma=-5^\circ$ (see Figure \ref{fig:Figure2}). Therefore the phase
shift of the modulated intensity peaks will be larger than $\phi_p$
(where the modulation peak is located), and this extra phase shift
becomes larger for the broader modulation. Furthermore, the intensity component
gets broader as well as shifted further in the negative $\sigma$ as the
modulation mapped on to broader region in observer's frame compared to
that in the positive $\sigma$ (see the case of $\sigma_\phi=0.1$ in
Figures \ref{fig:Figure5} and \ref{fig:Figure6}).

Circular polarization $V_s$ is antisymmetric for the steeper
modulation ($\sigma_\phi=0.1$), and changes sign from negative to
positive in the case of $\sigma=5^\circ$ and from positive to negative
in the case of $\sigma=-5^\circ$. In both cases
($\sigma=\pm5^\circ$), the positive circular becomes stronger compared to
the negative circular due to an enhancement of the unmodulated
positive circular over the negative circular (see Figure
\ref{fig:Figure3}). Also the phase location of the sign reversal of the
circular either leads or lags the intensity peak due to the above said
asymmetry. In the case of broader modulation ($\sigma_\phi=0.4$ and
$0.25$), the circular becomes almost positive throughout the pulse. Hence
it becomes almost symmetric as the effect of the modulation becomes
less significant over the beaming regions in comparison with the
induced asymmetry between the unmodulated positive and negative
circulars that resulted from PC-currents.

Linear polarization $L_s$ almost follows the total intensity $I_s$
profile with a lower magnitude due to an incoherent addition of emissions
within the beaming region. Further, it is slightly enhanced on the leading
side where the circular is smaller than on the trailing side in the case of
$\sigma=5^\circ,$ and vice versa in the case of $\sigma=-5^\circ$. PPA
is increasing (``counter clockwise'' or ``ccw'' swing as
$d\psi_s/d\phi'>0$) in the case of $\sigma=5^\circ$, while
decreasing (``clockwise'' or ``cw'' swing as $d\psi_s/d\phi'<0$) for
$\sigma=-5^\circ.$ Hence, in the case of pulsars with antisymmetric-type circular polarization we find the correlation of the sense reversal
of circular polarization from negative to positive with the increasing
PPA and vice versa. The inflection point of PPA is mostly unaffected
for the broader modulation, similar to the one with an unmodulated
emission (see Figure \ref{fig:Figure4}). But for a steeper modulation
($\sigma_\phi=0.1$), it is shifted to the earlier phase in the case of
$\sigma=5^\circ$ and to the later phase in the case of $\sigma=-5^\circ,$
which is opposite of the intensity phase shifts. Note that the phase
shift of the PPA inflection is found to be shifted toward the stronger
$L_s$ side. The phase shifts of the PPA inflection point in the cases of
$\sigma=\pm5^\circ$ with $\sigma_\phi=0.1$ are found to be
$0^\circ.73$ and $0^\circ.63,$ respectively. The smaller phase shift
of the PPA inflection point in the case of $\sigma=-5^\circ,$ as compared
to that in the case of $\sigma=5^\circ,$ is due to effective broadening of the 
modulation in the observer's frame.

For the next case we consider the modulation symmetrically located on
either side with respect to the meridional plane:
$\phi_p=180^\circ\pm30^\circ$ for $\sigma=-5^\circ$ and
$\phi_p=\pm40^\circ$ for $\sigma=5^\circ$. The simulated polarization
profiles are shown in Figure \ref{fig:Figure7}. The modulation
strength $f$ that the observer finds at the beaming region center and
the corresponding radius of curvature $\rho$ of the particle's trajectory
are also given. Even though $\rho$ is roughly symmetric at about
$\phi'=0,$ it becomes asymmetric in strength at the modulation peaks
between the leading and trailing sides. The $\rho$ at the modulation
peak on the leading side becomes smaller than that on the trailing side in
the case of $\sigma=-5^\circ$, and vice versa in the case of
$\sigma=5^\circ.$ This is due to the phase shift of the modulation
peaks (the mid point of the modulation peaks is indicated by the thick
arrows) from the phase shift of the emission points (see Figure
\ref{fig:Figure2}). As a result, an asymmetry arises in the strength of
the intensity components between the leading and the trailing sides: the component
on the leading side becomes stronger than that on the trailing side in the
case of $\sigma=-5^\circ$ and vice versa in the case of
$\sigma=5^\circ$. Also, since $\rho$ at the modulation peak on leading
side becomes less steep than that on the trailing side in the case
of $\sigma=-5^\circ$, the leading component becomes broader, and vice
versa in the case of $\sigma=5^\circ$ as indicated by the second row
panels. In both the cases of $\sigma$, similar to the modulation, the
outer phase emissions become stronger than the inner ones, and $V_s$
clearly shows that. However the larger asymmetry between the positive
and negative circulars under the leading side component in
the $\sigma=-5^\circ$ case and that under the trailing one in the $\sigma=5^\circ$ case is due to the selective enhancement of the positive
circular of the unmodulated emissions (see Figure \ref{fig:Figure3}).
\subsubsection{Emission with Polar Modulation}
By assuming a hallow cone modulation around the magnetic axis, i.e.,
a plasma density gradient only in the polar direction, we simulated the
polarization profiles, which are shown in Figures \ref{fig:Figure8} and
\ref{fig:Figure9}. The parameters used for the simulation are $f_0=1$,
$f_\phi=1$, and the rest are the same as in Figure
\ref{fig:Figure3}. In the cases of $\theta_p=3^\circ$ and
$\sigma_\theta=0.005$ and $0.002,$ since the minimum of
$\theta\sim(2/3)\sigma=3^\circ.3$ the observer sight line just grazes
the hallow emission cone and hence a single intensity component is
observed. But in the case of $\theta_p=4^\circ$ and
$\sigma_\theta=0.003$ the observer sight line cuts the hallow cone twice
and hence results in double components. In all cases, the intensity peak
(the $I_s$ peak in the cases of $\theta_p=3^\circ$ and
$\sigma_\theta=0.005,$ and $0.002$ and the mid point of the intensity
peaks in the case of $\theta_p=4^\circ$ and $\sigma_\theta=0.003$)
more or less remains at $\phi'=0^\circ$. This is because, even with
the perturbation of the dipole, both the colatitude and the
corresponding radius of curvature remain symmetric about
$\phi'=0^\circ.$ However there is a small phase shift of the PPA
inflection point in the cases of $\theta_p=3^\circ$ and
$\sigma_\theta=0.002,$ and $\theta_p=4^\circ$ and
$\sigma_\theta=0.003,$ but the shift is in the opposite direction
between the $\sigma=\pm5^\circ$ cases.

In the case of $\theta_p=3^\circ$ and $\sigma_\theta=0.005,$ a tiny
circular survives due to a less steep modulation, but in the cases
of $\theta_p=3^\circ$ and $\sigma_\theta=0.002,$ and
$\theta_p=4^\circ$ and $\sigma_\theta=0.003$ it is quite
significant. In the case of $\theta_p=3^\circ$ and
$\sigma_\theta=0.002$ where the observer sight line grazes the
modulation at a larger colatitude, a symmetric-type negative circular is
observed due to the selective enhancement of beaming region emissions
over the smaller $\theta$ part (see Figure \ref{fig:Figure3}). But in
the case of $\theta_p=4^\circ$ and $\sigma_\theta=0.003,$ an
antisymmetric-type circular is observed with the sign changing from negative
to positive under the leading component and vice versa on trailing. This
is due to the selective enhancement of emissions over either smaller
values of $\theta$ or larger depending upon when the sight line
crosses the hallow cone. Note that with the modulation in $\theta,$
the sign reversal of the circular is the same for both cases of $\sigma$
unlike the case with modulation in $\phi$ (see Figures
\ref{fig:Figure5} and \ref{fig:Figure6}). Since the beaming region
emission, due to the uniform source distribution, has a larger positive
circular than the negative circular (see Figure \ref{fig:Figure3}),
modulated profiles also show a slightly larger positive circular than
the negative circular.
\subsubsection{Emission with Modulation in both the Magnetic Colatitude 
and Azimuth}
In general, the plasma distribution can be nonuniform in both the
polar and azimuthal directions and hence we consider the modulation in
both $\theta$ and $\phi$. The extreme cases of modulation: modulation
in $\phi$ dominating over that in $\theta$ for
$\sigma_\phi\ll\sigma_\theta$ and the modulation in $\theta$
dominating over that in $\phi$ for $\sigma_\theta\ll\sigma_\phi,$ are
detailed in Sections $3.2.1.$ and $3.2.2.,$ respectively. The
simulations with the combinations of modulation parameters used in the
Sections $3.2.1.$ and $3.2.2.$ give the general polarization profiles.

For the case with the Gaussian modulation, with peaks located symmetrically
on either sides of the magnetic meridional plane in the conal ring of
colatitude $4^\circ$, we consider two sets of azimuthal peak locations
at $\phi_p=\pm50^\circ$ and $\pm30^\circ.$ The simulated polarization
profiles are given in Figure \ref{fig:Figure10} for the case of
$\sigma=5^\circ$. For simulation we used $f_0=1$,
$\sigma_\theta=0.01,$ $\sigma_\phi=0.1,$ and the other parameters are the
same as in Figure \ref{fig:Figure3}. Since the phase shift of the emission
points from the PC-current is only in $\phi$ and not in $\theta,$
the observer sight line selectively encounters the modulation depending
upon their azimuthal locations as shown in Figure
\ref{fig:Figure10}. The sight line encounters a stronger modulation on
the leading side as it passes closer to the Gaussian peak than that on the
trailing side in the case $\phi_p=\pm50^\circ,$ and vice versa in the
case $\phi_p=\pm30^\circ.$ However, the asymmetry in the strength of
the components between the leading and trailing sides becomes enhanced
or reduced in the intensity profiles depending upon the asymmetry in
the curvature at the modulation peaks as explained earlier for Figure
\ref{fig:Figure7}. The $\rho$ at the modulation peak on the leading
side becomes larger than that on the trailing side. Hence there is a decrease
in the asymmetry in the strengths of the modulated intensity
components between the leading and trailing sides in the case of
$\phi_p=\pm50^\circ,$ whereas it increases in the case of
$\phi_p=\pm30^\circ.$ The inner circulars become stronger than the
outer ones in the case of $\phi_p=\pm50^\circ$ as the observer
selectively encounters the modulated regions that are closer to the
magnetic meridional plane, and vice versa in the case of
$\phi_p=\pm30^\circ.$ Further, it is more pronounced on the trailing
side than that on the leading side due to an asymmetry in the modulation
encountered by the observer.

Note that if one considers the modulation in the $\theta,$ direction which is
even steeper, then the circular becomes an almost symmetric type with the
sign of inner circulars for $\phi_p=\pm50^\circ,$ and only the outer
circulars sign for $\phi_p=\pm30^\circ.$ Also note that if one chooses
negative sight line $\sigma=-5^\circ$ and the modulation peaks are
situated symmetrically in the azimuthal direction with respect to the
meridional plane, then the behavior of the strength of the modulation that
the sight line encounters and intensity components on the leading and
trailing sides will be opposite to the case of $\sigma=5^\circ.$
\subsection{Emission from the Southern Hemisphere} 
So far we have considered the emissions from the northern hemisphere of the
pulsar magnetosphere, which lies toward the positive $\mbox{\boldmath
  $\Omega$}.$ However one can receive emissions from the southern
hemisphere that lies toward the negative $\mbox{\boldmath $\Omega$}.$
By considering $\alpha=180^\circ-30^\circ$ and the azimuthal nonuniform
source distribution, we have simulated the polarization profiles for
the emission from the southern hemisphere for the two cases of
$\zeta=180^\circ-35^\circ$ and $180^\circ-25^\circ.$ The simulated
profiles are given in Figure \ref{fig:Figure11}. The simulations show
that the emission from the southern hemisphere for the case of
$\zeta=180^\circ-35^\circ$ is similar to the case of $\sigma_\phi=0.1$ of
Figure \ref{fig:Figure5}, except with the opposite sign of circular
polarization and opposite PPA swing. Similarly, the emission in the
case $\zeta=180^\circ-25^\circ$ is similar to that in the case
$\sigma_\phi=0.1$ of Figure \ref{fig:Figure6}. Note that this behavior
remains the same even with the nonuniform source distribution in both
the polar and azimuthal directions that we have considered in the case
of Figure \ref{fig:Figure10}.
\section{Discussion}
The PC-current-perturbed dipole field is axisymmetric with respect to the
magnetic axis wherein the field lines are twisted around the magnetic
axis due to an azimuthal component. Since the plasma particles (or
bunches) are constrained to move along the field lines, the beaming
direction of the relativistic emission gets aberrated toward the
direction of a tangent of twisted field lines. As a result, for the
case of positive $\sigma$ sight line, the emission points of the 
perturbed dipole field are shifted to later phases (or later times) in
the magnetic azimuth than those of the unperturbed dipole field, and
vice versa for the case of the negative $\sigma$ sight line. Since the
induced twist of field lines is around the magnetic axis only, the
emission points are found to be more or less unaffected in the
magnetic colatitude. As the perturbed dipole field remains
axisymmetric, for a given field line constant $r_e,$ the radius of the
curvature of the perturbed dipole field lines becomes a function of
colatitude only. Hence the curvature remains symmetric about
$\phi'=0,$ which is similar to the magnetic colatitude. Therefore
the emission due to the uniform radiation source distribution and its
polarization profiles remain symmetric about $\phi'=0$ as shown in
Figure \ref{fig:Figure4}, which is similar to the emission in the
unperturbed dipole field, except for the survival of a small circular
polarization due to a PC-current-induced asymmetry.

However, due to the nonuniform source distribution with a density
gradient effectively in the magnetic azimuth, the modulated intensity
components get shifted in the phase. Since the phase shift of the magnetic
azimuth between the two cases $(\pm\sigma)$ is of opposite
directions, the modulated intensity components shift to the later
phase in the case of positive $\sigma$ and to the earlier phase in the
case of negative $\sigma.$ Even though the radius of curvature $\rho$
is symmetric about $\phi'=0,$ it becomes asymmetric about the shifted
modulation peak where the observer encounters the maximum plasma
density. Hence the phase shift of modulated intensity components
becomes significantly different from that of the peak location of the
broader modulation. But the intensity components do not get shifted in
phase if the modulation is effective only in the polar direction as
both $\theta$ and $\rho$ remain symmetric about $\phi'=0.$

\citet{HA01} have shown that due to the distortion of the underlying
dipole field, the resultant PPA curve gets an upward shift in the PPA
versus rotation phase diagram, but the inflection point remains roughly
unchanged. Our simulations do confirm the results of \citet{HA01} in
the case of unmodulated emission. However, once the radiation source
distribution becomes nonuniform with larger density gradient, as we
showed in Sections $3.2.,$ the phase shift of the PPA inflection point
becomes quite significant. Note that, if there are phase shifts
between the intensity components and that of the PPA inflection point
then they will be in the opposite directions.

Since the phase shift of emission points in the perturbed dipole field
occurs only in the magnetic azimuth but not in the colatitude, the observer
in general encounters an asymmetry in the strength of the two
directional (polar and azimuthal) modulations between the leading and
trailing sides as shown in Figure \ref{fig:Figure10}. Depending upon
the azimuthal peak location of the Gaussian modulation in a given
cone, the sight line selectively encounters the modulation either on the
leading side or the trailing side. Hence, an asymmetry arises in the
strength of the intensity components between the leading and trailing
sides. However, the asymmetry in the intensity profiles could get
enhanced or reduced. This depends upon the asymmetry in the curvature
of the source trajectory at the modulation peaks between the two sides,
which is induced due to the phase shift of the modulation peaks. Note
that, even with modulation only in the azimuthal direction, an
asymmetry arises in the strength of the intensity components between the
leading and trailing sides as shown in Figure \ref{fig:Figure7}.

We showed that both the ``antisymmetric'' (with an asymmetry in the
strength between the opposite circular polarization polarities, in
general) and ``symmetric'' types of circular polarization could be
observed anywhere within the pulse window. This is possible due to an
induced asymmetry in the polarized emission from the beaming region,
such as the asymmetry between the positive and negative polarities, and
asymmetry due to the rotation of circular contour patterns in the
$(\theta,\phi)$-plane, the modulation of emission due to the
non-uniform distribution of sources.

Our simulations show that the sign reversal of circular polarization
near the central region of the pulse profile from negative to positive
is correlated with the increasing PPA (wherein the PPA curve has a
positive slope) and vice versa. This behavior is found in many pulsars
\citep{RR90} and has been confirmed by \citet{G10}, and
\citet{KG12}. However, there is also an association of the sign
reversal of the circular polarization from positive to negative with the
increasing PPA and vice versa \citep{Hanetal98,YH06}. In the case of
pulsars with the sign reversal of circular polarization not associated
with the central core region, \citet{YH06} have found that both the sign
reversals of the circular from negative to positive and from positive to
negative are associated with both the increasing and the decreasing PPA,
and our simulation of double component pulse profiles confirms this
behavior. In pulsars with the symmetric type of circular
polarization, we observed no correlation between the sense of circular
polarization and the PPA swing. Hence we confirm the findings of \citet{Hanetal98}
that the positive circular could be associated with either the
increasing or decreasing PPA, and similarly negative circular
too. Note that \citet{KG12} have also confirmed these findings in the
emissions from the rotating dipole.

The $\mbox{\boldmath $E$} \times \mbox{\boldmath $B$}$ drift on the
spark-associated plasma filaments makes them rotate around the
magnetic axis, and serves as a natural and physical mechanism for the
sub-pulse drift phenomenon \citep{RS75}. The effects of the
$\mbox{\boldmath $E$} \times \mbox{\boldmath $B$}$ drift and the
PC-current are similar, in the sense that both of them introduce an
azimuthal component of velocity to the plasma in addition to the velocity
in the direction of tangents to the dipolar field lines. The
aberration angle $\eta'$ \citep{G05} between the unperturbed velocity
$\mbox{\boldmath $v_0$}$ and the perturbed velocity $\mbox{\boldmath $v$},$
is given by $\cos\eta'=\hat v_0\cdot\hat{v},$ where $\hat
v_0=\mbox{\boldmath $v_0$}/|\mbox{\boldmath $v_0$}|$ and $\hat
v=\mbox{\boldmath $v$}/|\mbox{\boldmath $v$},$ can be used to
investigate the significance of one over the other. For the case of the
$\mbox{\boldmath $E$} \times \mbox{\boldmath $B$}$ drift, the
perturbed velocity is given by $\mbox{\boldmath $v$}=\mbox{\boldmath
  $v_0$} +\mbox{\boldmath $v_1$},$ where $\mbox{\boldmath $v_1$}=2 \pi
(\partial\mbox{\boldmath $r$}/\partial\phi)/\hat p_3$ is the drift
velocity around the magnetic axis with $\hat{p_3}$ being the time
taken by the plasma filament to complete one full rotation around the
magnetic axis. By considering observationally quoted values for
$\hat{p_3}$ in the literature \citep[see][and references
  therein]{GMG03}, we find the ratio of $\eta'$ due to $\mbox{\boldmath
  $E$} \times \mbox{\boldmath ${B}$}$ to that due to the
PC-current. It is $\sim3\%$-$4\%$ for slow drifting pulsars (e.g., it is
about $3\%$ for PSR B0943+10 and $4\%$ for PSR B2319+60 whose
$\hat{p_3}\sim37P$ and $\sim70P,$ respectively) and $\sim8\%$-$15\%$ for
fast drifting pulsars (e.g., it is about $8\%$ for PSR B2303+30 and
$15\%$ for PSR B0826$-$34 whose $\hat{p_3}\sim23P$ and $\sim14P,$
respectively). Therefore, we believe that the PC-currents are more
significant in comparison with the $\mbox{\boldmath $E$} \times
\mbox{\boldmath $B$}$ drift in influencing the emission beam geometry
and polarization, unless the pulsar drifting is very fast (i.e., $\hat
p_3\sim P$).

In this work, we considered the altitude-independent PC-current
density, but in reality it could have an altitude dependency due to
the diverging nature of the dipolar field lines and conservation of
escaping plasma. Since we considered constant emission altitude $r$
across the pulse profile, our simulations are still valid in
explaining the trend of the effect of the PC-current on pulsar
polarization profiles. However, if one considers the emissions from a
range of altitude that increases from inner to outer phases
\citep[e.g.,][]{GG01,GG03,DRH04}, then the altitude dependent current
density would be more appropriate. Our aim was to understand the
role of the PC-current on pulsar radio emission and polarization, and we did
not consider the pulsar rotation here. However, the combined role of
PC-currents and the rotation on the pulsar radio emission and
polarization needs to be investigated.
\section{Conclusion}
\begin{figure}
\centering
\epsscale{0.74}
\plotone{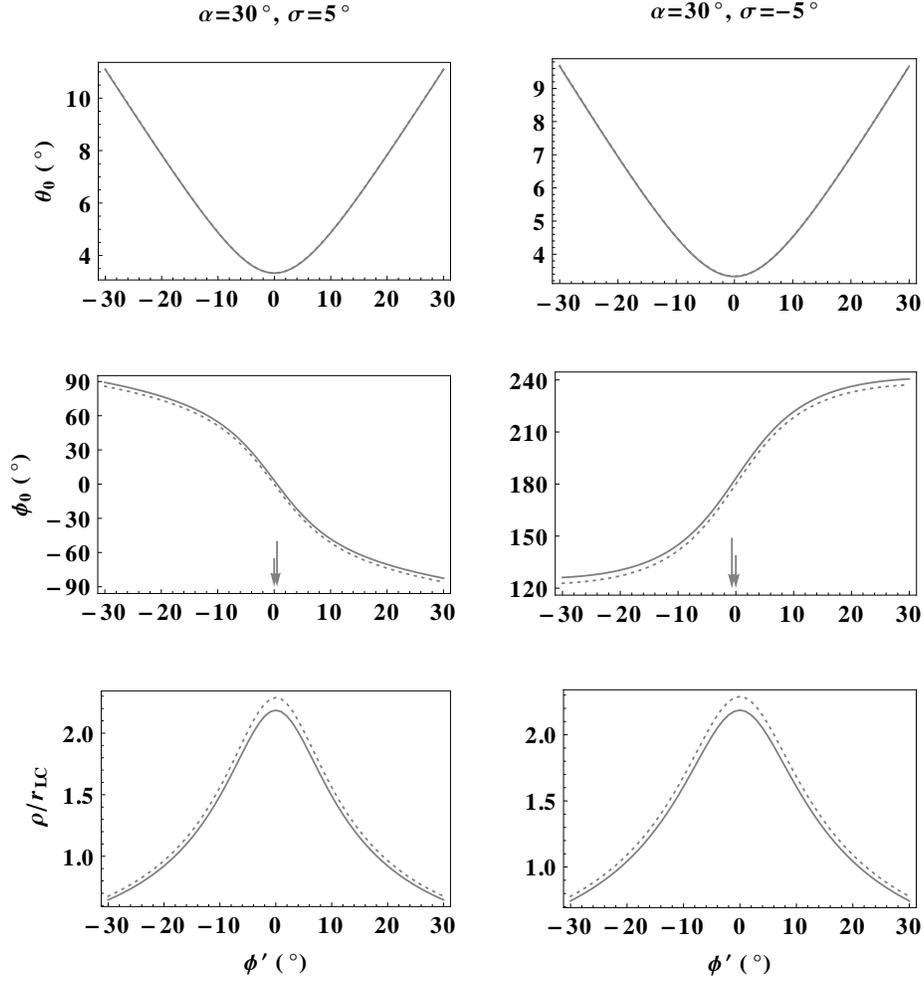}
\caption{Magnetic colatitude $\theta_{0}$ and azimuth $\phi_{0}$ of
  the beaming region center are plotted as functions of rotation phase
  $\phi'.$ The radius of curvature $\rho$ is plotted in the lower
  panels. The dotted and solid line curves are for the unperturbed and
  perturbed cases of the dipolar magnetic field. The parameters used are
  $r_{n}=r/r_{\rm LC}=0.1$, $P=1$ s, and $\varsigma=1.$ In the
  $\phi_{0}$ panels, the long and short arrows indicate the
  antisymmetric and inflection points of $\phi_0$ in the perturbed
  dipole, respectively.}
 \label{fig:Figure2}
\end{figure}

\begin{figure}
\centering
\epsscale{0.8}
\plotone{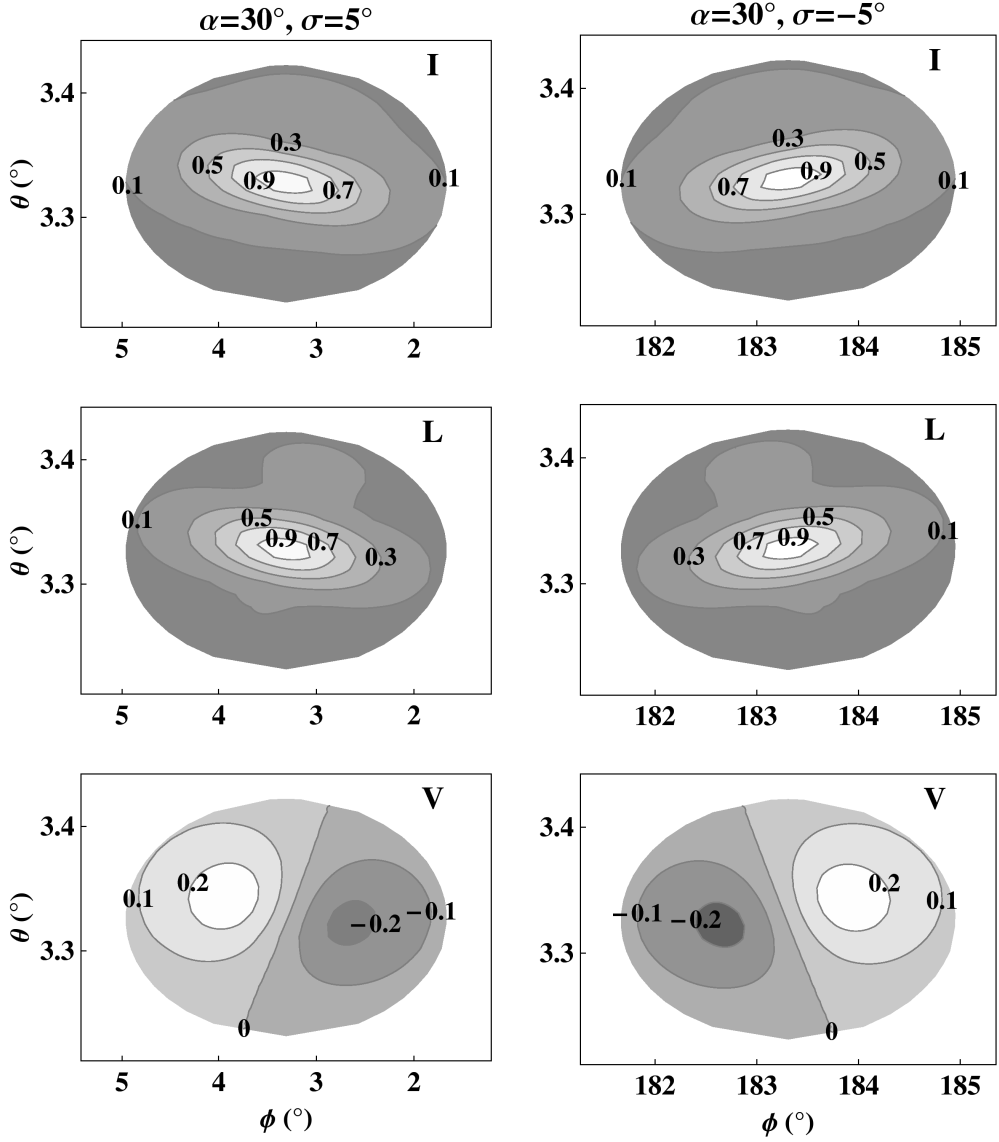}
\caption{Polarization state of emissions from the beaming regions
  with uniform radiation source density at rotation phase
  $\phi'=0^\circ$. For simulation we have used the parameters
  $\alpha=30^\circ$, $\sigma=\pm5^\circ$, $r_{n}=0.1$, $P=1$ s,
  $\varsigma=1$, $\gamma=400$, and $\nu=600$ MHz.}
 \label{fig:Figure3}
\end{figure}

\begin{figure}
\centering
\epsscale{.8}
\plotone{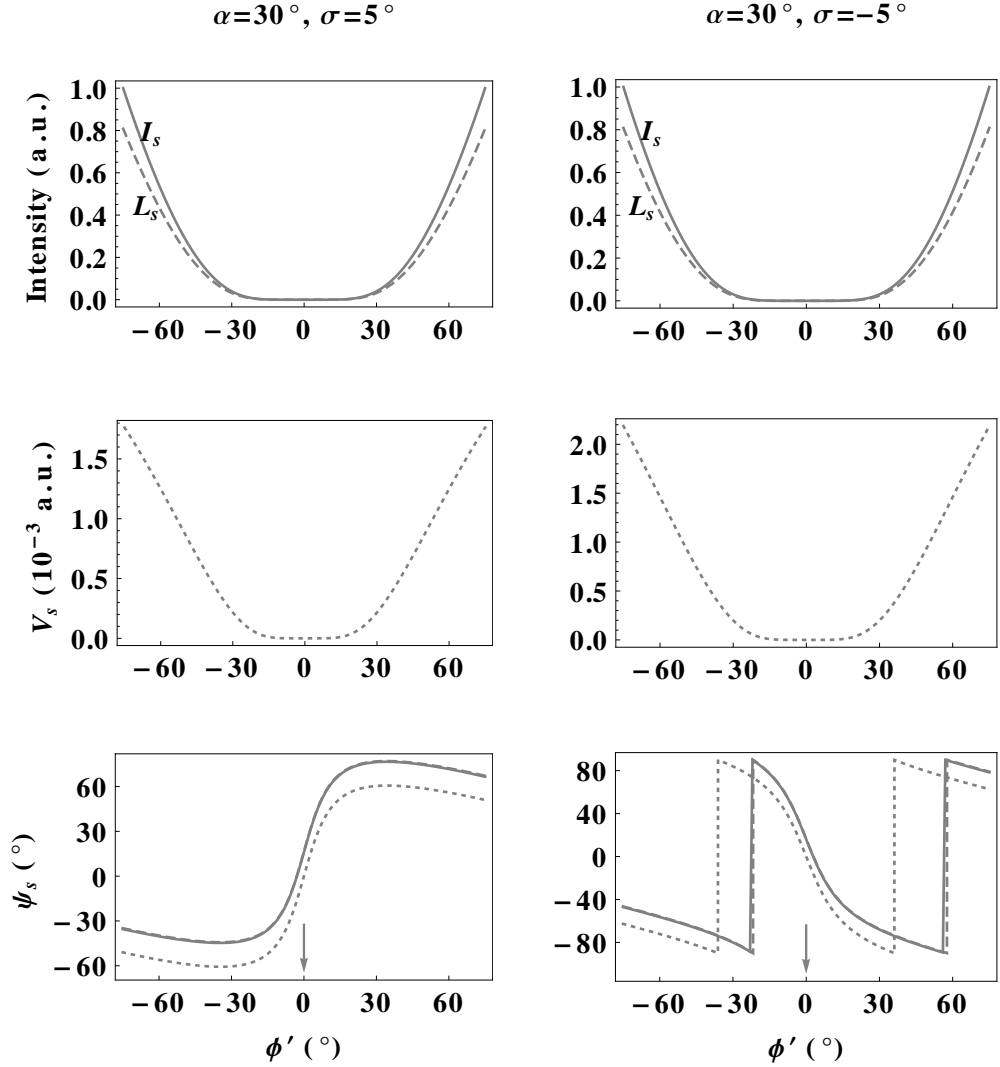}
\caption{Simulated polarization profiles for emission with uniform
  source distribution. In the PPA ($\psi_s$) panels, dotted line
  curves represent the RVM for the unperturbed dipole, whereas dashed
  and solid ones represent the PPA in the perturbed dipole geometry
  derived from \citet{HA01} and our calculations, respectively. The
  solid arrows point to the PPA inflection points from our
  simulation. The parameters chosen for simulation are the same as in
  Figure \ref{fig:Figure3}. }
 \label{fig:Figure4}
\end{figure}

\begin{figure}
\centering
\epsscale{0.9}
\plotone{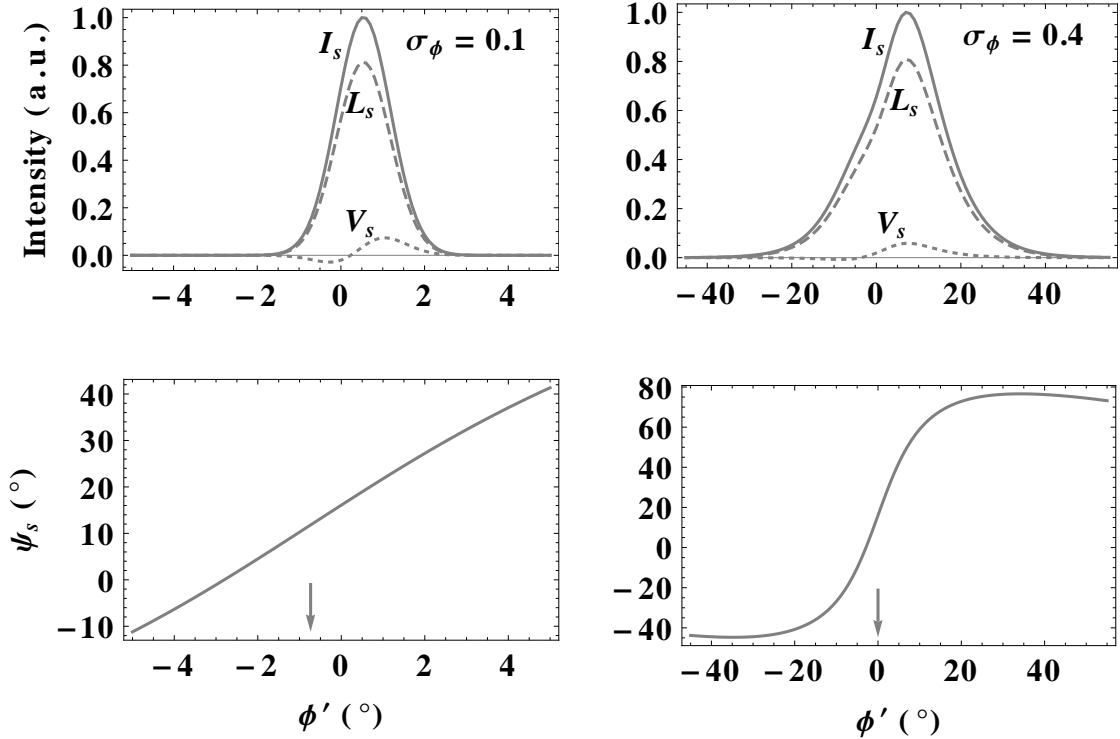}
\caption{Simulated polarization profiles for emission from the
  nonuniform distribution of sources (modulation) in azimuthal
  direction for the $\sigma=5^\circ$ case. The parameters used for
  simulation are $f_0=1$, $f_\theta=1$, $\phi_p=0^\circ,$ and the other
  parameters are the same as in Figure \ref{fig:Figure3}. }
 \label{fig:Figure5}
\end{figure}

\begin{figure}
\centering
\epsscale{0.9}
\plotone{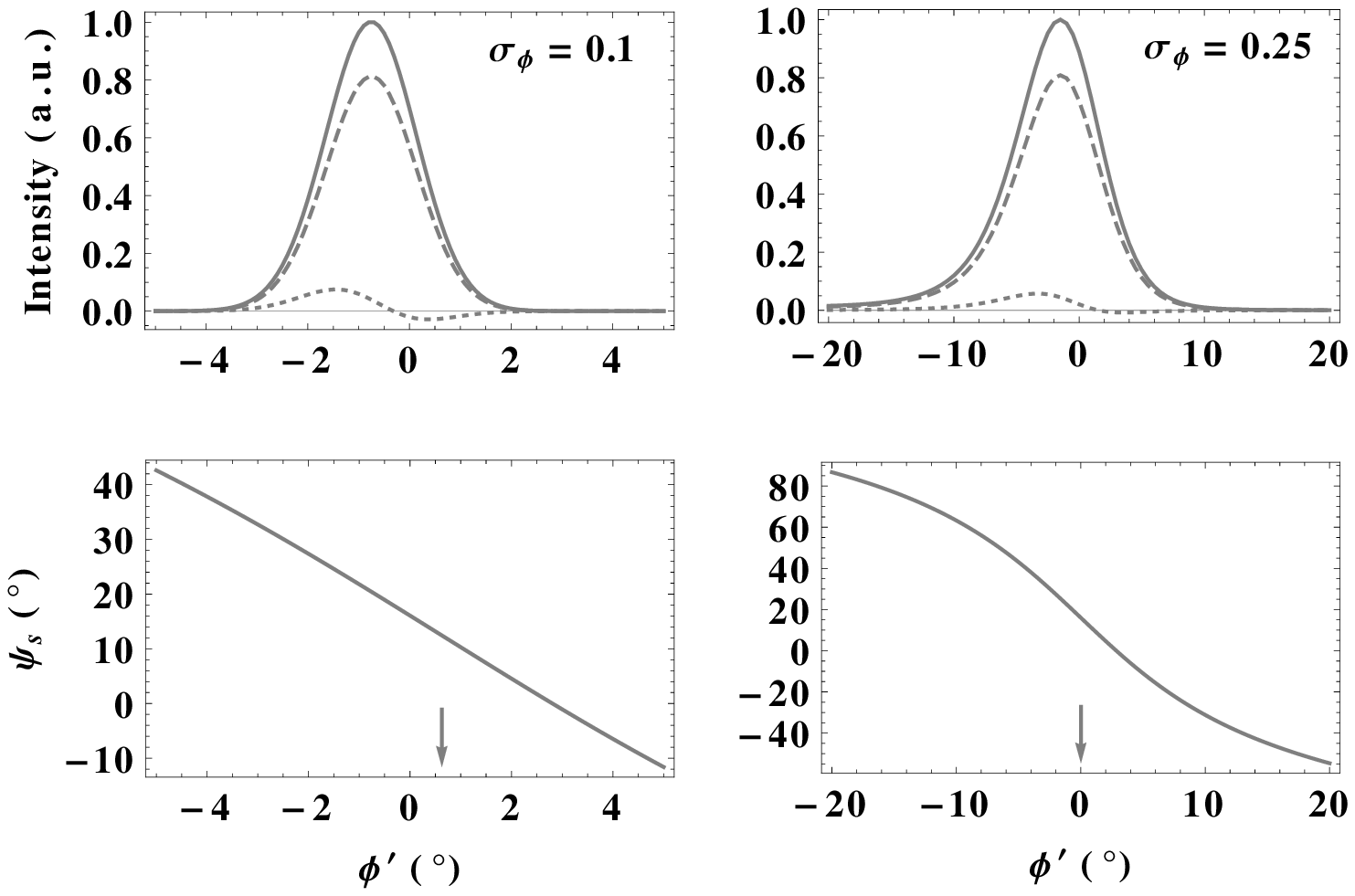}
\caption{Same as Figure \ref{fig:Figure5} except $\sigma=-5^\circ$ and
  $\phi_p=180^\circ$.}
 \label{fig:Figure6}
\end{figure}

\begin{figure}
\centering
\epsscale{0.8}
\plotone{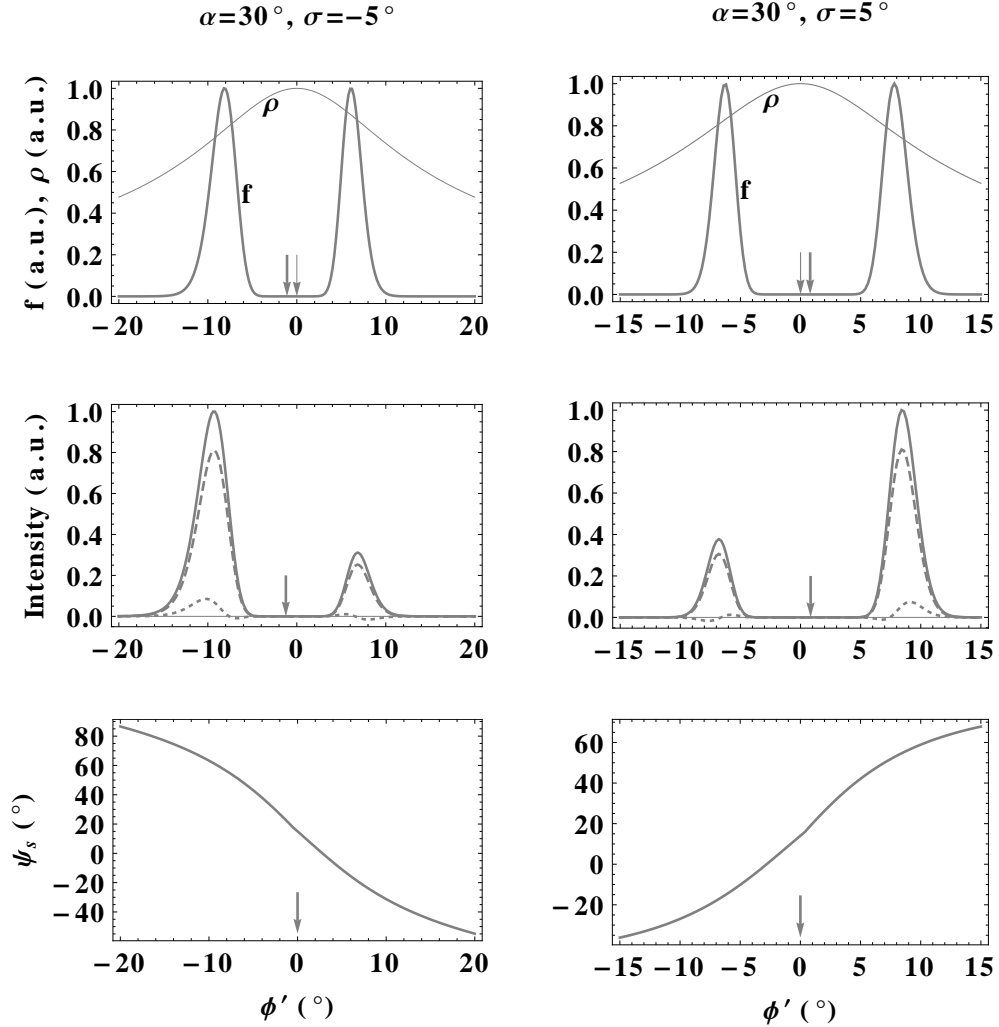}
\caption{Same as Figure \ref{fig:Figure5} except
  $\phi_p=180^\circ\pm30^\circ$ for $\sigma=-5^\circ,$ and $\phi_p=\pm
  40^\circ$ for $\sigma=5^\circ.$ In the top row panels, the thick
  line curve represents the modulation strength $f$ whereas the thin
  line curve represents the corresponding radius of curvature $\rho;$
  the thick line arrows mark the midpoint of the modulation peaks whereas
  the thin line arrows mark the symmetric point of $\rho$. Both $f$
  and $\rho$ are normalized with their corresponding maximum values.}
 \label{fig:Figure7}
\end{figure}

\begin{figure}
\centering
\epsscale{0.9}
\plotone{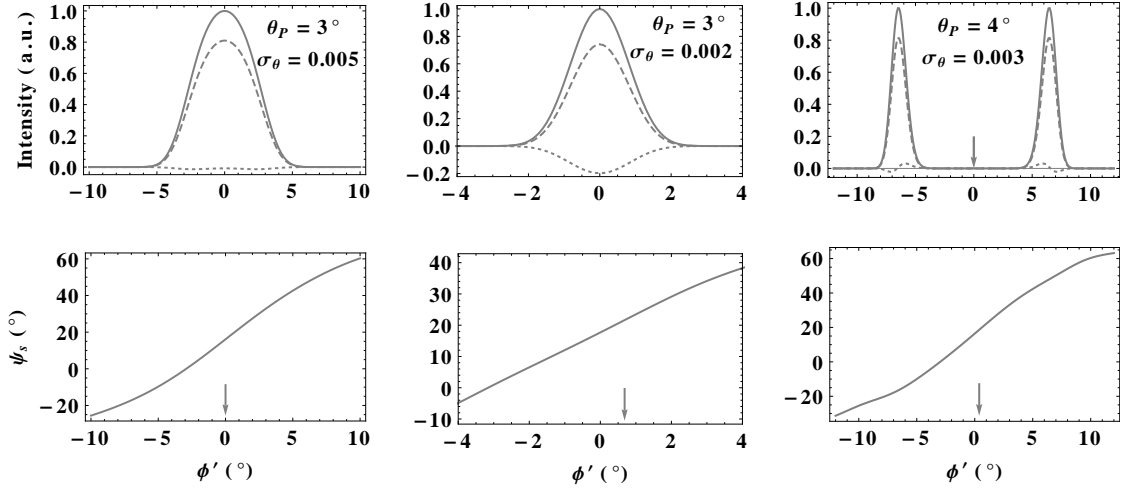}
\caption{Simulated polarization profiles for emission with a modulation
  in the polar direction for the case of $\sigma=5^\circ$. Here $f_0=1$,
  $f_\phi=1$, and the other parameters are the same as in Figure
  \ref{fig:Figure3}. }
 \label{fig:Figure8}
\end{figure}

\begin{figure}
\centering
\epsscale{0.9}
\plotone{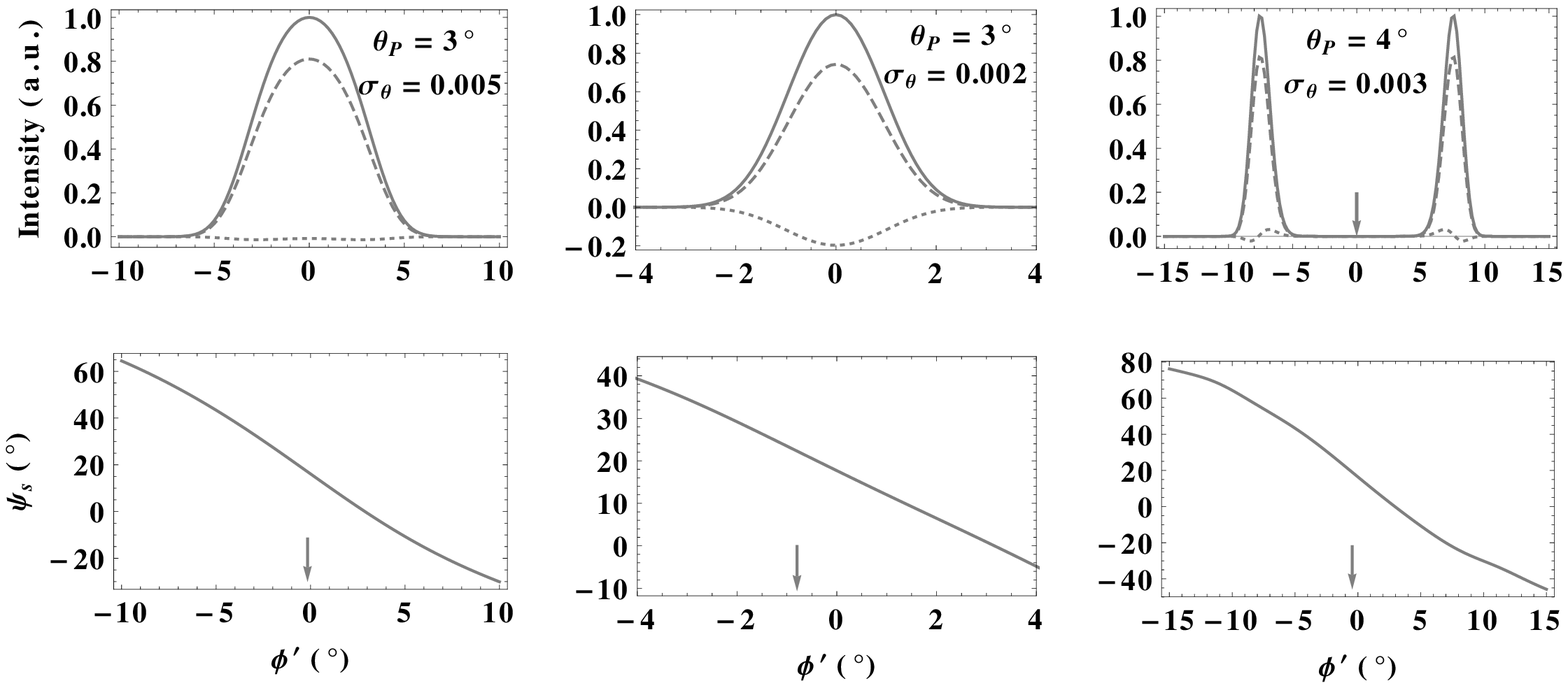}
\caption{Same as Figure \ref{fig:Figure8} except for $\sigma=-5^\circ$
  and $\phi_p=180^\circ.$}
 \label{fig:Figure9}
\end{figure}

\begin{figure}
\centering
\epsscale{0.9}
\plotone{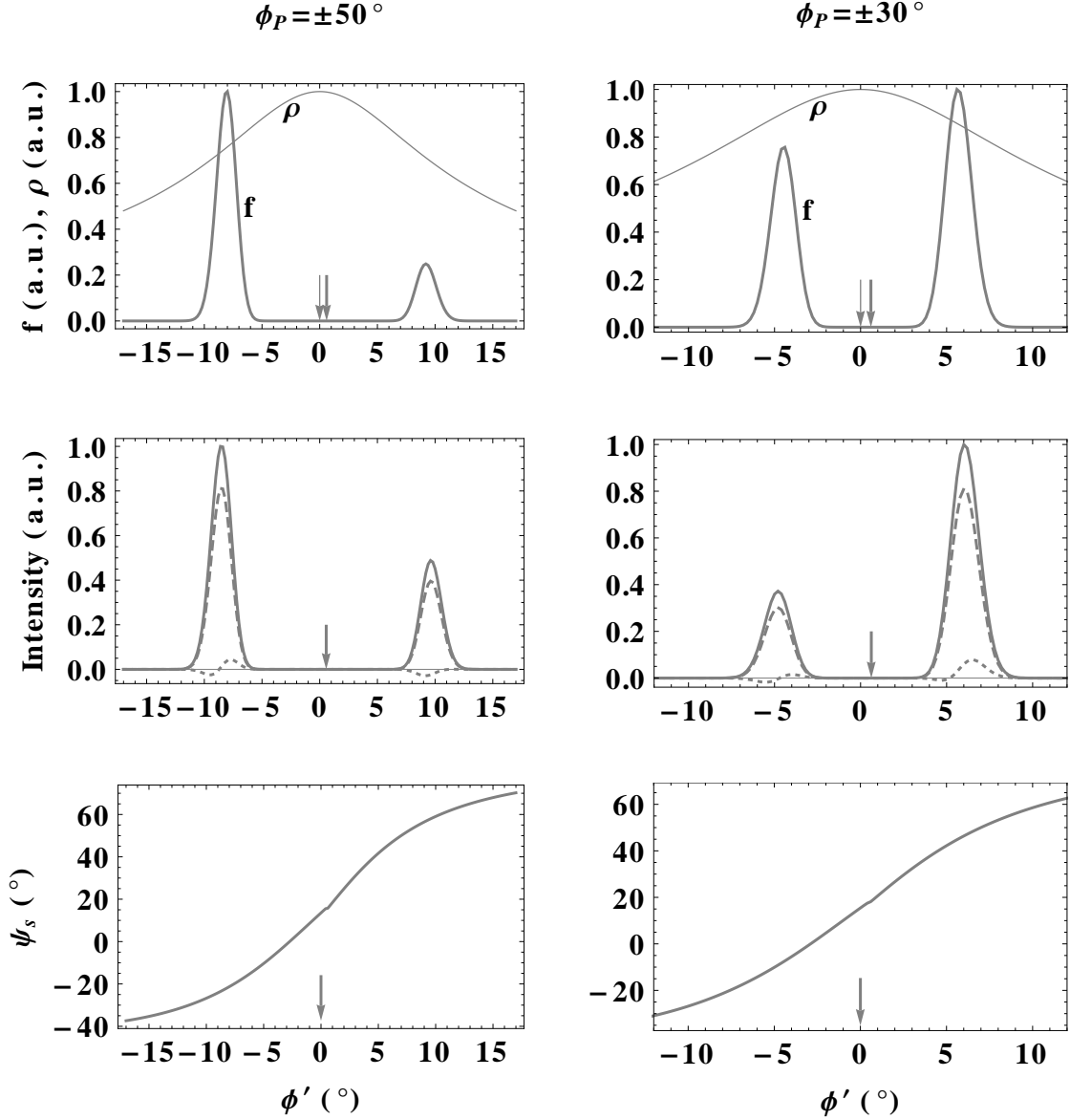}
\caption{Simulated polarization profiles for emission with a modulation
  in both the polar and azimuthal directions for the case of
  $\sigma=5^\circ$. Here $f_0=1$, $\theta_p=4^\circ,$
  $\sigma_\theta=0.01,$ $\sigma_\phi=0.1,$ and the other parameters
  are the same as in Figure \ref{fig:Figure3}. }
 \label{fig:Figure10}
\end{figure}

\begin{figure}
\centering
\epsscale{0.9}
\plotone{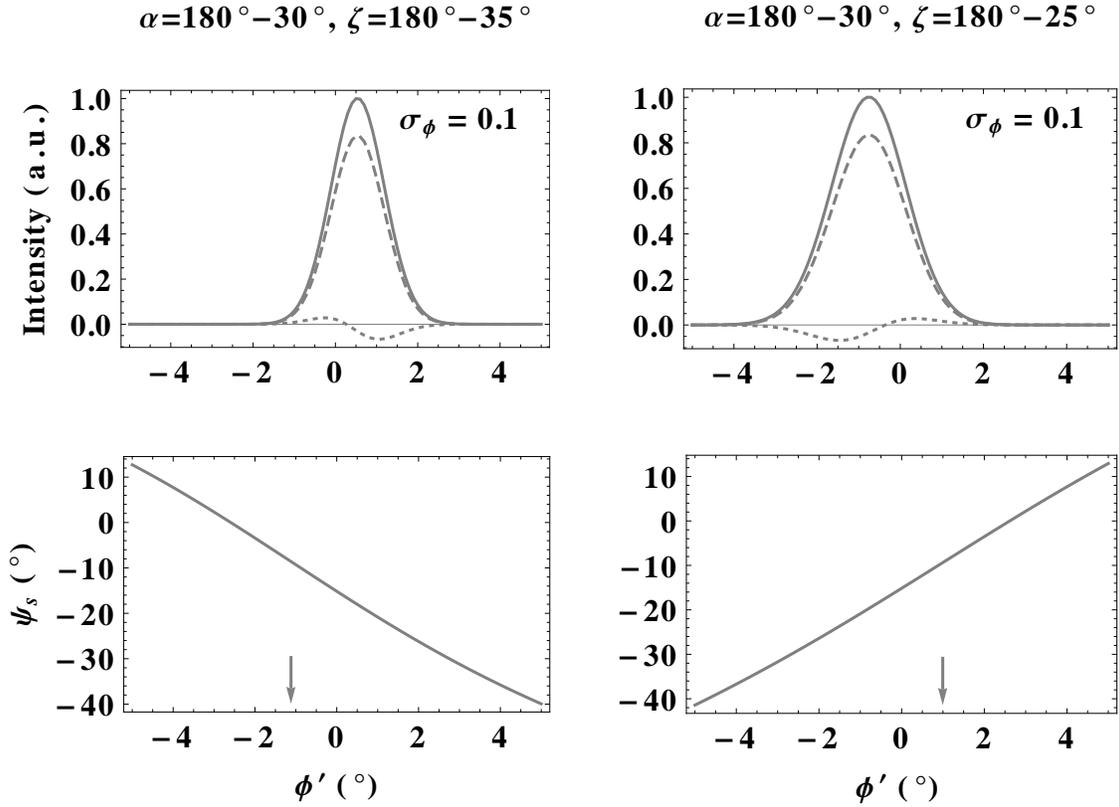}
\caption{Simulated polarization profiles for emission from the southern
  hemisphere for the cases of $\alpha=180^\circ-30^\circ$, and
  $\zeta=180^\circ-35^\circ$ and $180^\circ-25^\circ$. Here the other
  parameters used for simulation are the same as in Figures
  \ref{fig:Figure5} and \ref{fig:Figure6} except $\sigma=-5^\circ$ and
  $\phi_p=180^\circ$ for the $\zeta=180^\circ-35^\circ$ case, and
  $\sigma=5^\circ$ and $\phi_p=0^\circ$ for the $\zeta=180^\circ-25^\circ$
  case. }
 \label{fig:Figure11}
\end{figure}

By considering the PC-current-induced perturbation on the dipole field
we have developed a model for the pulsar radio emission and
polarization. By incorporating the nonuniform distribution of
radiation sources in the emission region of the distorted dipole
field, we have simulated a few typical polarization pulse
profiles. Based on our simulations, we conclude the following points.
\begin{enumerate}
\item Due to the PC-current-induced perturbation of the underlying
  dipole field, the emission points in the magnetic azimuth get shifted to
  later phases in the case of positive $\sigma,$ and to earlier phases
  in the case of negative $\sigma$ compared to those due to
  unperturbed ones (see Figure \ref{fig:Figure2}).
\item The intensity components that resulted from the modulation that
  dominated in the azimuthal direction over that in the polar direction, shift
  to either later or earlier phases depending upon the viewing
  geometry. The magnitude of the phase shift depends upon the steepness of the
  modulation and the PC-current-induced asymmetry in the curvature of the
  source trajectory.
\item Due to the PC-current-induced asymmetry and modulation, there
  arises a small phase shift in the PPA inflection point, too.
\item The leading side intensity components could become stronger and
  broader than the corresponding trailing ones or vice versa
  depending upon the viewing geometry and azimuthal location of the
  Gaussian modulation peaks in the concentric cones.
\item Both the ``antisymmetric''-type circular polarization with usual
  asymmetry in the strength of the opposite polarities and the
  ``symmetric''-type circular polarization become possible anywhere
  within the pulse window. It depends upon the viewing geometry and
  modulation in the emission region of the perturbed dipole field.
\end{enumerate}
We thank the anonymous referee for useful comments.

\clearpage  
\begin{thebibliography}{}
\bibitem[Ahmadi \& Gangadhara(2002)]{AG02} Ahmadi, P., \& Gangadhara,
  R.~T.\ 2002, \apj, 566, 365

\bibitem[Barnard \& Arons(1986)]{BA86} Barnard, J.~J., \& Arons,
  J.\ 1986, \apj, 302, 138

\bibitem[Blaskiewicz et~al.(1991)]{BCW91} Blaskiewicz, M., Cordes,
  J. M., \& Wasserman, I. 1991, \apj, 370, 643

\bibitem[Cheng \& Ruderman(1979)]{CR79} Cheng, A.~F., \& Ruderman,
  M.~A.\ 1979, \apj, 229, 348

\bibitem[Cheng \& Ruderman(1980)]{CR80} Cheng, A.~F., \& Ruderman,
  M.~A.\ 1980, \apj, 235, 576

\bibitem[Dyks(2008)]{D08} Dyks, J.\ 2008, \mnras, 391, 859 

\bibitem[Dyks et~al.(2004)]{DRH04} Dyks, J., Rudak, B., \&
  Harding, A.~K.\ 2004, \apj, 607, 939

\bibitem[Dyks et~al.(2010)]{DWD10} Dyks, J., Wright, 
G.~A.~E., \& Demorest, P.\ 2010, \mnras, 405, 509

\bibitem[Gangadhara(1997)]{G97} Gangadhara, R. T. 1997, A\&A, 327, 155

\bibitem[Gangadhara(2005)]{G05} Gangadhara, R.~T.\ 2005, \apj, 628, 923
 
\bibitem[Gangadhara(2010)]{G10} Gangadhara, R.~T.\ 2010, \apj, 710, 29

\bibitem[Gangadhara \& Gupta(2001)]{GG01} Gangadhara, R. T., \& Gupta,
  Y. 2001, \apj, 555, 31

\bibitem[Gil et al.(1993))]{Getal93} Gil, J. A., Kijak, J. \&
  Zycki, P. 1993, A\&A, 272, 207

\bibitem[Gil et al.(2003)]{GMG03} Gil, J., Melikidze, G.~I., \&
  Geppert, U.\ 2003, \aap, 407, 315

\bibitem[Gil \& Snakowski(1990a)]{GS90a} Gil, J. A., \& Snakowski,
  J. K. 1990a, A\&A, 234, 237

\bibitem[Gil \& Snakowski(1990b)]{GS90b} Gil, J. A., \& Snakowski,
  J. K. 1990b, A\&A, 234, 269

\bibitem[Goldreich \& Julian (1969)]{GJ69} Goldreich, P., \& Julian,
  W. H.\ 1969, \apj, 157, 869

\bibitem[Gupta \& Gangadhara(2003)]{GG03} Gupta, Y., \& Gangadhara,
  R. T. 2003, \apj, 584, 418

\bibitem[Han et~al.(1998)]{Hanetal98} Han, J. L., Manchester, R. N.,
  Xu, R. X., \& Qiao, G. J., 1998, \mnras, 300, 373

\bibitem[Hibschman \& Arons(2001)]{HA01} Hibschman, J.~A., \& Arons,
  J.\ 2001, \apj, 546, 382

\bibitem[Jackson(1998)]{Jackson98} Jackson, J. D. 1998, Classical
  Electrodynamics, (New York: Wiley)

\bibitem[Kramer et~al.(1994)]{Ketal94} Kramer, M., Wielebinski, R.,
  Jessner, A. , Gil, J. A. \& Seiradakis, J. H.  1994, \aaps, 107, 515

\bibitem[Kumar \& Gangadhara (2012)]{KG12} Kumar, D., \&
  Gangadhara, R. T. 2012, \apj, 746, 157


\bibitem[Lyne \& Manchester(1988)]{LM88} Lyne, A. G., \& Manchester,
  R. N. 1988, \mnras, 234, 477

\bibitem[McKinnon(1997)]{Mc97} McKinnon, M.~M.\ 1997, \apj, 475, 763

\bibitem[McKinnon \& Stinebring(1998)]{McS98} McKinnon, M.~M., \&
  Stinebring, D.~R.\ 1998, \apj, 502, 883

\bibitem[Melrose(2003)]{M03} Melrose, D.~B.\ 2003, in ASP
  Conf. Ser. 302, Radio Pulsars, ed. M. Bailes, D. J. Nice, \&
  S. E. Thorsett (San Francisco, CA: ASP), 179

\bibitem[Michel(1987)]{M87} Michel, F. C. 1987, \apj, 322, 822

\bibitem[Mitra \& Deshpande(1999)]{MD99} Mitra, D., \& Deshpande,
  A. A. 1999, \aap, 346, 906

\bibitem[Mitra \& Rankin(2002)]{MR02} Mitra, D., \& Rankin,
  J.~M.\ 2002, \apj, 577, 322

\bibitem[Radhakrishnan \& Cooke(1969)]{RC69} Radhakrishnan, V., \&
  Cooke, D. J. 1969, ApJ, 3, L225

\bibitem[Radhakrishnan \& Rankin(1990)]{RR90} Radhakrishnan, V., \&
  Rankin, J. M. 1990, ApJ, 352, 258

\bibitem[Rankin(1983)]{R83} Rankin, J. M. 1983, \apj, 274, 333

\bibitem[Rankin(1990)]{R90} Rankin, J. M. 1990, \apj,  352, 247

\bibitem[Rankin(1993)]{R93} Rankin, J. M. 1993, \apjs, 85, 145

\bibitem[Ruderman \& Sutherland(1975)]{RS75} Ruderman, M. A., \&
  Sutherland, P. G. 1975, ApJ, 196, 51

\bibitem[Sturrock(1971)]{S71} Sturrock, P. A.  1971, ApJ, 164, 229

\bibitem[Thomas \& Gangadhara(2007)]{TG07} Thomas, R.~M.~C., \&
  Gangadhara, R.~T.\ 2007, \aap, 467, 911

\bibitem[Thomas \& Gangadhara(2010)]{TG10} Thomas, R.~M.~C., \&
  Gangadhara, R.~T.\ 2010, A\&A, 515, 86

\bibitem[Thomas et~al.(2010)]{Tetal10} Thomas, R.~M.~C.,
  Gupta, Y., \& Gangadhara, R.~T.\ 2010, \mnras, 406, 1029

\bibitem[Xilouris et al.(1998)]{Xetal98} Xilouris, K.~M., 
Kramer, M., Jessner, A., et al.\ 1998, \apj, 501, 286 


\bibitem[You \& Han(2006)]{YH06} You, X. P., \& Han, J. L. 2006,
  Chin. J. Astron. Astrophys., 6, 237

\end{thebibliography}
\end{document}